\documentclass[aps,a4paper,showpacs,prc,floatfix,onecolumn,amsmath,amssymb,nofootinbib]{revtex4-1}
\usepackage{graphicx}
\usepackage{epsfig}
\usepackage{ulem}
\usepackage{morefloats}
\usepackage{rotating}
\usepackage{xcolor,ulem}
\usepackage{graphicx}

\begin{document}

\title{Electrical Conductivity of Hydrogen Plasmas: Low-density Benchmarks and Virial Expansion Including $e-e$ Collisions}

 \author{G. R\"{o}pke}
 \email{gerd.roepke@uni-rostock.de}
 \affiliation{Institut f\"{u}r Physik, Universit\"{a}t Rostock, D-18051 Rostock, Germany}
\date{\today}

\begin{abstract}
An improved virial expansion for the low-density limit of the electrical conductivity $\sigma (T,n)$ of hydrogen as the simplest ionic plasma is presented. 
Quantum statistical methods provide exact values for the lowest virial coefficients, which serve as a benchmark for analytical approaches to electrical conductivity as well as for numerical results from density functional theory based molecular dynamics simulations (DFT-MD) or path-integral Monte Carlo (PIMC) simulations. 
The correction  factor introduced by Reinholz {\it et al.}, Phys. Rev. E {\bf 91}, 043105 (2015) is applied to describe the inclusion of electron-electron collisions in DFT based calculations of transport coefficients. As a benchmark, the first virial coefficient is correctly described with this approach. The value of the second virial coefficient is discussed, questions about its value according to DFT-MD simulations are addressed.  
\end{abstract}

\maketitle

\section{Introduction}

Coulomb plasmas such as the hydrogen plasma are strongly interacting many-particle systems in which the formation of bound states (atoms and their ionisation stages) is possible. 
Depending on the parameter values temperature $T$ and particle number density $n_e=n_i=n$ (we consider a charge-neutral hydrogen plasma consisting of electrons $e$, mass $m_e$, and protons $i$, mass $m_i$), the electrons can be bound or degenerate, so that a quantum statistical approach is required. Other ionic plasmas with atomic nuclei, charge number $Z$, can be treated in a similar way, but are not considered here. Instead of the variables $T$ and $n$, which characterise the state of the plasma, the dimensionless plasma parameter 
\begin{equation}
\Gamma = \frac{e^2}{4\pi\epsilon_0 k_{\rm B}T} \left(\frac{4\pi}{3} n\right)^{1/3}
\end{equation}
is usually introduced as the ratio of potential to kinetic energy in the non-degenerate case, and the electron degeneracy parameter 
\begin{equation}
\Theta = \frac{2m_e k_{\rm B}T}{\hbar^2} (3\pi^2 n)^{-2/3}.
\end{equation}

The dc conductivity $\sigma_{\rm dc}(T,n)$ of the hydrogen plasma was calculated in the low-density limit in the seminal work by Spitzer and H\"arm \cite{Spitzer53} using kinetic theory. The Boltzmann equation valid in the low-density limit and the corresponding Fokker-Planck equation were solved, with the result
\begin{equation}
\sigma^{\rm Spitzer}(T,n)=\frac{(k_{\rm B}T)^{3/2} (4\pi\epsilon_0)^2}{m_e^{1/2} e^2}\frac{1}{0.84624 \ln (\Theta/\Gamma)}.
\end{equation}

In order to treat arbitrary densities, the linear response theory was worked out, which considers the plasma under the influence of a weak external field.
Properties of the plasma are determined by correlation functions. According to linear response theory, the dc conductivity is related to equilibrium fluctuations of the charge current (which is mainly the electron current due to the large mass ratio $m_i/m_e \gg 1$), resulting in the Kubo formula \cite{Kubo66}
\begin{equation} 
\label{Kubo}
\sigma(T,n) = \frac{e^2 \beta}{3 m_e^2\Omega}\int_{-\infty}^0 dt e^{-i z t}\int_0^1 d \lambda \,{\rm Tr}[{\bf P \cdot P}(t+i \hbar \beta\lambda) \rho_0]
\end{equation} 
where $\beta=1/k_{\rm B}T$, $\Omega$ is the volume, $\bf P$ the total momentum of electrons, the $t$ dependence is according to the Heisenberg picture 
\begin{equation}
 O(t) = e^{i Ht/\hbar} O e^{-i Ht/\hbar},
 \end{equation}
 and
\begin{equation}
\rho_0= e^{-\beta(H-\sum_c\mu_cN_c)}/{\rm Tr}e^{-\beta(H-\sum_c\mu_cN_c)}
 \end{equation}
is the equilibrium statistical operator. Finally  the limit $z=i \eta \to 0$ has to be taken.

Equilibrium correlation functions can be evaluated using the methods of quantum statistics such as the Green's function method \cite{FW,GRBuch}.
This is a perturbative approach in which an expansion with respect to the Coulomb interaction is performed  \cite{Roep88,RR89,Heidi}.
  By applying partial summations describing, e.g., quasiparticles, screening and  the formation of bound states, useful results are obtained for a wide range of plasma parameters. The application to electrical conductivity was also shown in \cite{Rcond2018
 }. For the relation to kinetic theory see \cite{Reinholz12}.
 However,  exact results are only obtained in special limiting cases.
 In the following, we discuss the virial expansion of the inverse conductivity of the hydrogen plasma, which contains these exact results. They can be employed as benchmarks for any approach.

In order to avoid perturbation theory for the electron-ion interaction, the method of DFT-MD simulations was worked out \cite{Desjarlais}. 
The ion dynamics is treated classically using molecular dynamics simulations, while the electrons are treated as quantum particles, with the wave equations being solved using the DFT formalism. 
The evaluation of the correlation function (\ref{Kubo}) is performed numerically using the Kubo-Greenwood formula.
 However, the treatment of the electron-electron ($e-e$) interaction is approximated by a suitable choice of the exchange-correlation functional $E_{xc}$ \cite{Perdew1996}. This problem is also a subject of the present work.  
 
The rigorous treatment of the $e-e$ interaction is 
possible with the help of  path-integral Monte Carlo (PIMC) simulations, see \cite{Mil01,Bo20,Boehme22} and references therein.
However, the calculations are very complex and the current shortcomings of this approach include the relatively small number of particles and the sign problem for fermions.

In this work, we focus on two aspects: the virial expansion of the resistivity of hydrogen plasmas in the low-density region, and how DFT-MD simulations can be improved to describe the contribution of  electron-electron collisions to plasma conductivity. In Sec. \ref{sec:2} we discuss the virial expansion of the resistivity. The DFT-MD simulations of plasma conductivity are presented in Sec. \ref{sec:DFT}, and ways to account for $e - e$ collisions are discussed in Sec. \ref{sec:e-e}. Finally, we draw conclusions in Sec. \ref{sec:Concl}.

\section{Benchmarks from the virial expansion of the resistivity}
\label{sec:2}. 
\subsection{The standard virial expansion} 

For the sake of simplicity, we consider hydrogen plasmas. The results can be transferred to other plasmas with ionic charge $Ze$, since the low-density limit considered here is dominated by the Coulomb force and short-range interactions contribute to higher orders of the virial expansion of the resistivity.

To investigate the conductivity of the hydrogen plasma, we introduce  the dimensionless conductivity $\sigma^*(T,n)$ according to
\begin{eqnarray}
 \sigma(T,n) &=& \frac{(k_{\rm B}T)^{3/2} (4\pi\epsilon_0)^2}{m_e^{1/2} e^2}\;\sigma^*(T,n) 
= \frac{32405.4}{\Omega {\rm m}}T_{\rm eV}^{3/2} \sigma^*(T,n) \,,
 \label{eq:1}
\end{eqnarray}
where $T_{\rm eV}=k_{\rm B}T/{\rm eV}=11604.6 \,T/{\rm K}$ denotes the temperature measured in units eV.

Using the generalised linear response theory, a virial expansion was proposed for the dimensionless resistivity $\rho^*(T,n)=1/\sigma^*(T,n)$  \cite{Roep88,RR89,Redmer97}
\begin{eqnarray}
 \rho^*(T,n)= \rho_1 (T) \ln \left(\frac{\Theta}{\Gamma}\right) + \rho_2 (T) + {\cal O}\left[n^{1/2} \ln \left(\frac{\Theta}{\Gamma}\right) \right]
 \label{eq:virial}
\end{eqnarray}
with ($T_{\rm Ha}=27.21137\, T_{\rm eV}, n_{\rm Bohr}=n\, a_{\rm Bohr}^3$)
\begin{equation}
\frac{\Theta}{\Gamma}=\frac{1}{ (36 \pi^5)^{1/3}}\frac{2m_e}{\hbar^2} \frac{4 \pi \epsilon_0}{e^2} \frac{(k_{\rm B}T)^2}{n}=\frac{T_{\rm Ha}^2}{4n_{\rm Bohr}} (96 \pi^5)^{-1/3} \propto \frac{T^2}{n}\,.
\label{GamT}
\end{equation}
It describes the well-known logarithmic density dependence, the Coulomb logarithm, which is caused by the long-range Coulomb interaction.

According to Spitzer and H\"arm  \cite{Spitzer53}, who calculated the value $\gamma_E=0.5816$, the first virial coefficient 
\begin{equation}
\label{eq:Spitzer}
 \rho_1(T) =\rho^{\rm Spitzer}_1=\frac{\pi^{3/2}}{2^{7/2} \gamma_E} = 0.84624
\end{equation}
 is not dependent on $T$. For the second virial coefficient, the high-temperature limit follows from the Quantum Lenard-Balescu (QLB) approach
  \cite{RR89,RRT89,Redmer97,Karachtanov16}
\begin{equation}
 \lim_{T \to \infty}\rho_2(T) =  \rho_2^{\rm QLB} = 0.4917~.
 \label{virLB}
\end{equation}
In this limit the conductivity of hydrogen plasmas is approximated as
\begin{equation}
\label{sigmavir}
\sigma(T,n)=\frac{32405.4 \,T_{\rm eV}^{3/2}}{0.84624 \ln(\Theta/\Gamma)+0.4917}\frac{1}{\Omega {\rm m}}.
\end{equation}

The temperature dependence of $\rho_2(T)$ is not exactly known, an approximation can be obtained from the interpolation formula \cite{RRT89,Esser03}
\begin{equation}
 \rho_2(T) \approx 0.4917 + 0.846 \ln\left[ 
 \frac{1 + 8.492/T_{\rm eV}}{1 + 25.83/T_{\rm eV} + 167.2/T_{\rm eV}^2} \right]\,.
 \label{WKB}
\end{equation}
 It is a challenge for numerical simulations such as the PIMC simulation to improve this approximation for the temperature-dependent second virial coefficient.

The virial expansion serves as a benchmark for the theoretical and experimental determination of the plasma conductivity in the low-density range. To extract the virial coefficients from data for $\sigma(T,n)$, a virial plot  was proposed in \cite{RSRB21}. To determine $\rho_1(T)$ and $\rho_2(T)$, we draw 
\begin{equation}
\label{rho1eff}
\rho_1^{\rm eff}(T,n)=\rho^*(T,n)\frac{1}{\ln(\Theta/\Gamma)}=\frac{1}{\sigma^*(T,n)}\frac{1}{\ln(\Theta/\Gamma)}
\end{equation}
as a function of $x=1/\ln(\Theta/\Gamma)$.  Since
\begin{equation}
\rho_1^{\rm eff}(T,n)=\rho_1(T)+\rho_2(T) x+ {\cal O}[n^{1/2}]
\end{equation}
we expect a linear relationship in the range of low densities, where the ordinate at $x=0$ is $\rho_1$ and the slope is $\rho_2$. 
To perform the extrapolation to $x=0$, the variable  $x$ must be sufficiently small for linear behavior to be observed.

This method was applied to several approaches for determining the electrical conductivity of hydrogen plasmas, 
in particular to the data presented in \cite{Grabowski}. 
As shown in \cite{RSRB21}, several semi-empirical expressions for plasma conductivity fail to fulfill this benchmark.

The next term in the virial expansion (\ref{eq:virial}), which is of the order $n^{1/2}\ln(\Theta/\Gamma)$, is called the Debye-Onsager relaxation effect. An approximate expression for this third virial term was given in Ref. \cite{Roep88}. This higher order virial coefficient can be extracted by analysing the effective second virial coefficient 
 \begin{equation}
 \rho_2^{\rm eff}(T,n)= \rho^*(T,n)-\rho_1\ln(\Theta/\Gamma)=\rho_2(T)+\rho_3(T)n^{1/2}\ln(\Theta/\Gamma)+ {\cal O}\left(n^{1/2}\right)
 \end{equation}
as a function of $x_1=n^{1/2}\ln(\Theta/\Gamma)$. 
The corresponding virial plot gives $ \rho_2(T)$ as the value of $ \rho_2^{\rm eff}(T,n)$ at $x_1=0$, 
while $\rho_3(T)$ follows from the slope at $x_1=0$. 
The third virial term in the approximation of Ref. \cite{Roep88} is only a small correction and is not considered here. 
To extract it from a virial plot, accurate simulations or measurements of the plasma conductivity are required.

\begin{table}[htp]
\caption{Electrical conductivity $\sigma(T,n)$ of hydrogen plasmas from virial expansion, Eq. (\ref{sigmavir}), $\sigma_{\Lambda}$ from Eq. (\ref{sigmaLambda}), $\sigma_{\hat \Lambda}$ from Eq. (\ref{sigmahatLambda}). The density $\rho=1$ g cm$^{-3}$ corresponds to the particle number density $n=5.98 \times 10^{23}$ cm$^{-3}$, the density $\rho=1.67$ g cm$^{-3}$ to  $n=1 \times 10^{24}$ cm$^{-3}$, and $\rho=10$ g cm$^{-3}$ to $n=5.98 \times 10^{24}$ cm$^{-3}$.}
\begin{center}
\begin{tabular}{|c|c|c|c|c|c|c|c|c|c|c|c|}
\hline
$\rho$ (g cm$^{-3}$) 	& $T$ (eV)	& $\Gamma $	& $\Theta $	& $1/\ln(\Theta/\Gamma)$	& $\sigma$ ($\Omega^{-1}$ cm$^{-1}$)&
$\sigma_{\Lambda}$ ($\Omega^{-1}$ cm$^{-1}$)& $1/\Lambda$ &$\sigma_{\hat \Lambda}$ ($\Omega^{-1}$ cm$^{-1}$)& $1/{\hat \Lambda}$ \\ 
\hline
 1		& 2000	& 0.0097773	& 77.284	& 0.1114	& 3.58498$\times 10^6$ 	&  3.58497$\times 10^6$ & 0.1003 & 3.58496$\times 10^6$ & 0.1046 \\
 1		& 1000	& 0.0195546	& 38.642	& 0.1318	& 1.48255$\times 10^6$ 	&  1.48252$\times 10^6$ & 0.1164 &  1.48251$\times 10^6$ & 0.1224 \\
 1		&700		& 0.0279351	& 27.049	& 0.1454	& 951335 				& 951229 				& 0.127 	   &  951281 			& 0.1341 \\
 1		&400		& 0.0488865	& 15.457	& 0.1737	& 483512 				&  483446				& 0.1481 &  483412 			& 0.1578 \\
 1		&200		& 0.097773	& 7.7284	& 0.2288	& 218811 				&  218659				& 0.1862 &  218582 			& 0.2018 \\
 1		&100		& 0.195546	& 3.8642	& 0.3352	& 107445				&  107035				& 0.2504 &  106826 			& 0.2789 \\
 1		&70		& 0.279351	& 2.7049	& 0.4405	& 78668.2				&  77910.1			& 0.3035 &  77531 			& 0.3456 \\
\hline
 1.67	& 20000		& 0.00116		& 549.05	& 0.07653	&7.93757$\times 10^7$ 	&7.93757$\times 10^7$ 	& 0.0711 &7.93757$\times 10^7$ & 0.07327\\
 1.67	& 10000		& 0.00232		& 274.53	& 0.08561	&3.12362$\times 10^7$ 	& 3.12362$\times 10^7$ 	& 0.07888 &3.12362$\times 10^7$ & 0.08155\\
 1.67	& 1000		& 0.0232		& 27.453	&0.1413	&1.58184$\times 10^6$ 	& 1.58179$\times 10^6$	& 0.1239 &1.58176$\times 10^6$	& 0.1306\\
 1.67	& 100		& 0.232		& 2.7453	&0.4047	& 125498				& 124570				&0.2865 & 124103			&0.324\\
 10		& 1000	& 0.0421291	& 8.3252	& 0.1892	& 2.06434$\times 10^6$ 	& 2.06386$\times 10^6$ 	& 0.1591 &2.06361$\times 10^6$ & 0.1704 \\
 10		& 100	& 0.421291	& 0.83252	& 1.468	& 303434				&274960				&0.5529 &263134				&0.687\\
 100	& 1000		& 0.0907644	& 1.7936	&0.3352	& 3.39771$\times 10^6$ 	& 3.38473$\times 10^6$ 	& 0.2504 &3.37813$\times 10^6$ & 0.2789 \\
\hline
\end{tabular}
\end{center}
\label{Tab:1}
\end{table}%

\subsection{Generalized virial expansions}

The occurrence of a logarithmic term $\propto \ln(1/n)$ in the virial expansion of the resistivity, Eq. (\ref{eq:virial}), requires some discussion. We need to define a reference density $n^{\rm ref}(T)$ to make the logarithm dimensionless, $\ln(1/n) \to \ln(n^{\rm ref}(T)/n)$. In Eq. (\ref{eq:virial}), $n^{\rm ref}(T)=n \Theta/\Gamma $ was taken. 
The choice of the reference density influences the higher virial coefficients. 
In principle, $\hat n^{\rm ref}(T)$ can be chosen such that the next virial coefficient disappears, $\hat \rho_2(T)=0$, as shown below.

Another problem  is that the logarithm can become zero or negative, $\ln(n^{\rm ref}/n)\le 0$ for $n^{\rm ref}/n\le 1$, but the conductivity or resistivity are positively defined quantities. 
Therefore, it is of interest to perform a modified virial expansion using the Coulomb logarithm $\Lambda(T,n)$ from kinetic theory
(for references see \cite{French22}),
\begin{eqnarray}
 \rho^*(T,n)= \rho_1 \Lambda(T,n) +\tilde  \rho_2 (T) + {\cal O}[n^{1/2} \Lambda(T,n)],
 \label{eq:virC}
\end{eqnarray}

\begin{equation}
\label{Lamb}
\Lambda(T,n)=\ln(1+b)-\frac{b}{1+b},
\end{equation}
where
\begin{equation}
\label{Bornb}
 b(T,n)=\frac{3 (k_{\rm B}T)^2}{\pi n} \, \frac{4 \pi \epsilon_0 m}{e^2 \hbar^2}=\frac{n_\Lambda(T)}{n} =\frac{3^{4/3} \pi^{2/3}}{2^{1/3}} \frac{\Theta}{\Gamma}=\frac{3}{\pi}  \frac{T_{\rm Ha}^2}{n_{\rm Bohr}}
\end{equation}
is the Born parameter $\Theta/\Gamma$ (up to a factor). We have in the low-density limit
\begin{equation}
\lim_{n \to 0}\Lambda(T,n)=\ln[b(T,n)]-1.
\end{equation}
 The first virial coefficient $\rho_1(T)$ remains unchanged, 
but the higher virial coefficients are modified if $\ln(n^{\rm ref}/n)$ with  $n^{\rm ref}=n \Theta/\Gamma $ is replaced by $\Lambda(T,n)$ in all terms of the virial expansion. In particular, from Eq. (\ref{eq:virC}) we obtain 
\begin{equation}
\lim_{T \to \infty}\tilde \rho_2(T)=\lim_{T \to \infty}\rho_2(T)+\rho_1\left[1-\frac{1}{3} \ln(81 \pi^2/2)\right] = 0.4917+0.84624 [1-\frac{1}{3} \ln(81 \pi^2/2)]=-0.351938.
\end{equation}
The conductivity follows as 
\begin{equation}
\label{sigmaLambda}
\sigma_\Lambda(T,n)=\frac{32405.4 \,T_{\rm eV}^{3/2}}{0.84624\, \Lambda-0.351938}\frac{1}{\Omega {\rm m}}
\end{equation}

The value of $b(T,n)$, Eq. (\ref{Bornb}), is related to the collision integral evaluated for a screened potential, see \cite{ GRBuch}, but changes when the static screening is replaced by dynamic screening. 
We can scale $b(T,n)$ by a factor $y$ without changing the analytical structure of the virial expansion.
It is possible to set the second virial coefficient to zero at $T \to \infty$ if the parameter $b$ is replaced by $\hat b=b/y$ with $y=1.5158$. 
Then we have 
\begin{equation}
 \rho^*(T,n)= \rho_1 \hat \Lambda(T,n) +\hat  \rho_2 (T) + {\cal O}[n^{1/2} \hat \Lambda(T,n)],
 \label{eq:virChat} 
\end{equation}
where $\hat \Lambda(T,n)$ follows from Eq. (\ref{Lamb}) replacing $b$ by $\hat b$. 
We have $\lim_{T \to \infty} \hat  \rho_2 (T) =0$, and in the high-temperature limit
\begin{equation}
\label{sigmahatLambda}
\sigma_{\hat \Lambda}(T,n)=\frac{32405.4 \,T_{\rm eV}^{3/2}}{0.84624 \,\hat \Lambda}\frac{1}{\Omega {\rm m}}.
\end{equation}
Note that a $T$-dependent $y(T)$ can be introduced to compensate for the second virial coefficient for all $T$. These modifications of the virial expansion lead to changes in the higher virial coefficients. It is a question to the higher order expansions whether such modifications would result in a better convergence.

As example, the values for the conductivity for specific values of $T,n$ are shown in Tab. \ref{Tab:1}.
The values for low densities and high $T$ are considered where the parameter $x,\tilde x=1/\Lambda$ or $\hat x=1/{\hat \Lambda}$ is small, so that the virial expansion with error bars below few percent is valid. The largest discrepancy occurs for the relatively high density 10 g/cm$^3$ and the low temperature 100 eV.
The method of virial plots  can also be used to extract the virial coefficients $\rho_1(T), \rho_2(T), \tilde \rho_2(T)$  or $ \hat \rho_2(T)$ from experimental data or numerical simulations. An example is discussed in Sec. \ref{sec:DFT} below, see also \cite{RSRB21}.  

The method of virial plots has also been applied to thermodynamic variables describing the equation of state of plasmas, see \cite{RDVBB24}. 
There, higher-order virial coefficients and generalized virial expansions are discussed. 
Higher-order virial coefficients are of interest to describe degeneracy as known from the Lee-More model, see, e.g., \cite{French22}.

\section{The Kubo-Greenwood formula and DFT-MD simulations}
\label{sec:DFT}

The DFT-MD approach has been successfully applied to calculate the thermodynamic properties of complex materials in a wide range of $T$ and $n$. 
In the Born-Oppenheimer approximation, the ions are treated as classical particles whose time-dependent configuration is described by molecular dynamics simulations.
They act like an external potential on the electron subsystem. 
Together with the kinetic energy and Coulomb's $e-e$ interaction, a many-electron Schr\"odinger equation is formulated, which determines the many-body wave functions.
In density functional theory,  the correlated many-electron wave function is replaced by the antisymmetrized product of optimized  single-electron states, the Kohn-Sham orbitals.
For the corresponding  Kohn-Sham equations, the Hamiltonian contains the kinetic energy of a non-interacting reference system, the mean-field Coulomb interaction energy, and an exchange-correlation energy $E_{xc}$ that accounts to some approximation for the effects of $e-e$ correlations and antisymmetrization.  
A simple approximation uses expressions for $E_{xc}$  that are derived for the uniform electron gas.
%
For frequency-dependent electrical conductivity, see (\ref{Kubo}) for $z=\omega +i \eta$, the Kubo-Greenwood formula \cite{Kubo66,Greenwood}
\begin{equation}
\label{eq:KG}
\operatorname{Re} \left[\sigma (\omega)\right] = \frac{2 \pi e^2}{3 m_e^2 \omega \Omega} \sum_k w_k \sum_{j=1}^N \sum_{i=1}^N  \sum_{\alpha=1}^3 
\big[ f(\epsilon_{j,k})- f(\epsilon_{i,k}) \big]  |\langle{\Psi_{j,k}| \hat{p}_\alpha | \Psi_{i, k}}\rangle|^2 \delta(\epsilon_{i,k} -\epsilon_{j,k} - \hbar\omega )
\end{equation}
was used to calculate the frequency-dependent dynamic electrical conductivity $\sigma(\omega)$ in the long-wavelength limit~\cite{RSRB21,Desjarlais02,Mazevet05,Holst11,French2017,Gajdos2006}. Kohn-Sham wave functions $\Psi_{i,k}$ from density functional theory calculations are used to calculate the transition matrix elements of the electron momentum operator $\hat{p}_\alpha$, $\alpha =\{x,y,z\}$. The Fermi-Dirac distribution $f(\epsilon)$ accounts for the average occupation at energy $\epsilon$, and the summation over momentum space $k$ contains the $k$-point weights $w_k$. 


Due to the finite size of the simulation box, the $\delta$-function in equation~\eqref{eq:KG} must be approximated by a Gaussian with finite width, which also prevents the direct calculation of the DC conductivity at $\omega = 0$. Therefore, the dynamic conductivity is extrapolated to the limit $\omega \rightarrow 0$ using a Drude fit.
The extrapolation procedure to $\omega = 0$ can be improved by using a frequency-dependent collision frequency \cite{Heidi}.
%

One of the main shortcomings of the DFT-MD approach is that the many-particle interaction is replaced by a mean-field potential.  When using product wave functions for the many-electron system, correlations are excluded. The exchange-correlation energy density functional reflects the Coulomb interaction in  a certain approximation, e.g., as it exists in the homogeneous electron gas.
However, it becomes problematic in the low-density limit, where correlations, especially  $e-e$ collisions, are important.  

DFT-MD simulations have been successfully used to calculate the transport properties in the warm-dense matter region for various materials. 
The transition to condensed matter, the liquid or solid state, is adequately described, in particular for the degenerate electron system.
The question whether the contribution of $e-e$ collisions to the conductivity is correctly described also in the low-density range  has long been the subject of controversial debate.
The $e-e$ Coulomb interaction is included in the exchange-correlation energy of the DFT functional so that it was argued that the "ab initio" approach also describes the Coulomb interaction of the electrons \cite{Desjarlais,Starrett20}. 
However, it was pointed out \cite{Heidi} that DFT-MD is not able to reproduce the Spitzer limiting value of plasma conductivity, what leads to significant deviations in the low-density range.
The correct behavior of the plasma conductivity at low density was determined using generalized linear response theory (gLRT).

\begin{figure}[t]
\centerline{\includegraphics[width=0.7 \textwidth]{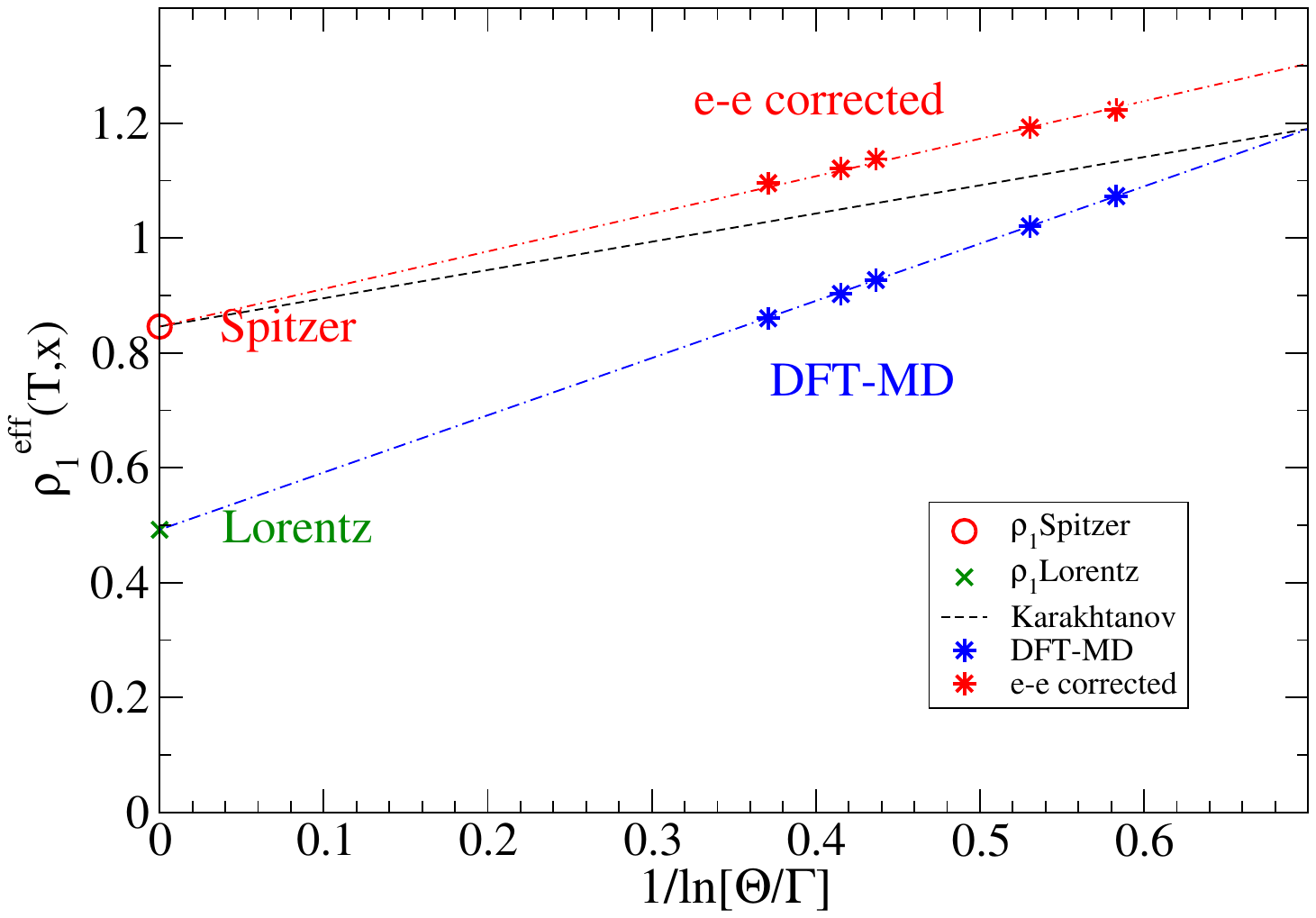}}
\caption{Reduced resistivity $\rho_1^{\rm eff}(T,x)$, Eq.  (\ref{rho1eff}), for hydrogen plasma as a function of $x=1/\ln (\Theta/\Gamma)$: 
DFT-MD simulations from Ref. \cite{RSRB21} and the generalized linear response result of Redmer and Karakhtanov~\cite{RR89,RRT89,Redmer97,Karachtanov16} in the high-temperature limit.  
$\rho^{\rm Spitzer}_1=0.846$, Eq. (\ref{eq:Spitzer}),  and $\rho^{\rm Lorentz}_1=0.492$, Eq. (\ref{Lorentz}), are defined in the text. The dot-dashed lines are linear relations connecting the Spitzer or Lorentz limit with the respective reduced resistivity value for the smallest $x$ value.
}
\label{Fig:DFTeecorr1}
\end{figure}

The method of virial plots was used to show that the DFT-MD simulations fail to describe the correct low density behavior of the electrical conductivity of the hydrogen plasma \cite{RSRB21,SCCS22}, see Fig. \ref{Fig:DFTeecorr1}. Instead of the correct virial expansion (\ref{eq:virial}) for the resistivity of the hydrogen plasma,
\begin{equation}
\label{VirH}
 \rho^*(T,n) \approx 0.84624 \ln \left(\frac{\Theta}{\Gamma}\right) +0.4917
\end{equation}
at high temperatures, the DFT-MD simulations are extrapolated as
\begin{equation}
\label{VirDFT}
 \rho^{*,\rm DFT-MD}(T,n) \approx 0.492 \ln \left(\frac{\Theta}{\Gamma}\right) +0.9886.
\end{equation}
This is the virial expansion for the Lorentz plasma, a model plasma where the Hamiltonian contains only the electron-ion interaction, the electron-electron interaction is replaced by the interaction within the homogeneously charged jellium to compensate the Coulomb divergency of the ion system,
\begin{equation}
H^{\rm Lorentz} =\sum_{c=i,e,k} \frac{p_{c,k}^2}{2 m_c}+\frac{1}{2}\sum_{k \ne l} V'_{ii}({\bf r}_{i,l} - {\bf r}_{i,k})+\sum_{j,k} V'_{ie}({\bf r}_{e,l} - {\bf r}_{i,k})
 \label{LorentzH}
\end{equation}
($V'$ means the exclusion of the Hartree term in the Coulombic pseudopotential interaction (screened by electrons)  due to the charge neutrality by the negative background) which has the first virial coefficient 
\begin{equation}
\rho_1^{\rm Lorentz} =\frac{1}{16} (2\pi^3)^{1/2}=0.492126\,.
 \label{Lorentz}
\end{equation}
The second virial coefficient $\lim_{T \to\infty}\rho_2^{\rm DFT-MD}(T)=0.9886$ is determined by the Born approximation, but includes the screening.  In \cite{SCCS22} it was discussed that in addition to the screening of the ion subsystem, which is given by the pair distribution function resulting from the MD simulations, the screening by the electrons according to the exchange-correlation energy density functional is also included. However, dynamical screening is not described by $E_{xc}$, in contrast to the quantum Lenard-Balescu equation.

We would like to mention that in the case of thermal conductivity it has been shown that 
the contribution of $e-e$ collisions is not taken into account in DFT-MD simulations \cite{Starrett20,Desjarlais,French22}
and results in an additional term. Other approaches such as generalized linear response theory can be used to solve this problem. 

The DFT-MD simulations have been used very successfully to calculate the transport properties of warm and dense matter (WDM). Electron-electron collisions are not treated adequately.  For degenerate systems, $e-e$ collisions are suppressed by Pauli blocking, so that their contribution can be neglected and the results of DFT-MD are applicable.
However, the systematic inclusion of $e-e$ collisions  is indispensable for the description of plasmas at $\Theta > 1$.

\section{The account of electron-electron collisions in numerical simulations of plasma conductivity}
\label{sec:e-e}

\subsection{The correction factor for Lorentz plasmas}

We refer to  a Lorentz plasma as a model system of electrons and ions, in which the interaction of electrons with ions is considered, see Eq. (\ref{LorentzH}).
The electron-electron interaction is only considered in the mean-field approximation, so that the electron-ion interaction is screened, 
but the effect of electron-electron collisions on the transport coefficients is neglected. 
For this Lorentz model plasma, the solution of the linearized Boltzmann equation is found using the relaxation time approach.
The relaxation time ansatz is possible for elastic collisions, which is the case for $\lim (m_i/m_e) \to \infty$, but not for $e-e$ collisions.

For the Lorentz model plasma, the electrical conductivity $\sigma^{\rm Lorentz}(T,n)$ can be calculated beyond perturbation theory.
The effective interaction of electrons with the ion subsystem is treated, for example, using the density functional theory. Similar approaches are given by the average atom model or the Lee-More model. These approaches are applicable in the high-density, degenerate case ($\Theta \le 1$), in which electron-electron collisions are suppressed due to Pauli blocking. 
They fail to describe real plasmas such as hydrogen plasmas in the low-density range, as $e-e$ collisions are not taken into account.

We introduce a correction factor 
\begin{equation}
\label{reduction}
R_{ee}(T,n)=\frac{\sigma(T,n)}{\sigma^{\rm Lorentz}(T,n)}
\end{equation}
which relates the plasma conductivity $\sigma(T,n)$ to the conductivity of the Lorentz model plasma. 
This correction factor $R_{ee}(T,\Theta)$ (we replace the density parameter $n$ by $\Theta$) approaches 1 for $\Theta \le 1$ and the value
$0.591 \pi^{3/2}/2^{5/2}= 0.582$ for  $\Theta \gg 1$.

Exact expressions for the conductivity $\sigma(T,n)$ of plasmas are obtained from the generalized linear response theory (gLRT) in the form of correlation functions. The kinetic theory (KT) follows as a special case \cite{Reinholz12}. However, the analytical evaluation of the correlation functions is only possible in some approximations using Green function diagram techniques. 
For example, the dynamic screening of the electrons is treated in the random phase approximation (RPA),  
the ionic structure factor is also approximated, the Born approximation can be improved by a T-matrix approach.

An expression for $R_{ee}(T,\Theta)$ was given in Ref. \cite{Heidi}, where both $\sigma(T,n)$ and $\sigma^{\rm Lorentz}(T,n)$ are treated by the same approximation (gLRT with two moments, Born approximation, RPA). 
It is near to 1 in the region of degeneracy $\Theta \le 1$ and approaches the limiting value 0.581 for $\Theta \to \infty$. 
It is calculated numerically by solving some integrals, see \cite{Heidi}.
For arbitrary ionic charges $Z$ a smooth interpolation formula was given, which reads for hydrogen ($Z=1$) in the low-density, classical limit (see Eq. (C.20) of Ref.  \cite{Heidi})
\begin{align}
\label{eq:ReeFit}
\lim_{T \to \infty}  R_{ee}(T,\Theta)= 0.594456+0.575578 \frac{1}{ \ln\left(0.0033305\,\Theta^{3/2}\,T_{\rm K}^{1/2}\right)} ,
\end{align}
where $T_{\rm K}=11604.6\, T_{\rm eV}$ is the temperature measured in Kelvin. 
In the context of the virial expansion, using $x=1/\ln (\Theta/\Gamma)$, we obtain 
\begin{equation}
\lim_{T \to \infty}R_{ee}(T,x)= 0.594456+0.575578\, x
\label{Reevir}
\end{equation}
 in lowest order of $x$.

Within the framework of a refined calculation, we must investigate the effects 
 of dynamical screening, ion-ion structure factor, and strong collisions, see Refs.~\cite{RR89,Redmer97,Reinholz05,Karachtanov13}. These effects are of minor importance for the correction due to $e-e$ collisions, 
both in the high-density limit where $\lim_{\Theta \to 0}R_{ee}(\Theta) = 1$ and
in the low-density limit, where they only occur in higher orders of the virial expansion. 
Although these corrections are small, we analyze them in the next section \ref{Example}.

\subsection{Example: DFT-MD simulations for hydrogen plasmas}
\label{Example}

As shown in Fig. \ref{Fig:DFTeecorr1}, the DFT-MD simulations for the effective resistivity $\rho_1^{\rm eff, DFT-MD}$ are on a straight line pointing to the Lorentz limit, see Ref. \cite{RSRB21}. The corrected  values $\rho_1^{\rm eff, ee-corr}=\rho_1^{\rm eff, DFT-MD}/R_{ee}$ are shown in Tab. \ref{tab:Max1}. The correction factor $R_{ee}$, Eq. (\ref{eq:ReeFit}), leads to a considerable increase (up to a factor $\approx 2$ in the low-density range) of the resistivity and solves the drawback of the correct $\rho_1$ benchmark by construction. 
However, the slope determining the second virial coefficient  seems to remain too large compared to the high-temperature limit (\ref{virLB}) obtained from QLB or gLRT approaches \cite{Redmer97,Karachtanov16}. 
While  the slope of the DFT-MD simulations is 0.9965, the slope of the $e-e$ corrected virial line is 0.732. The slope of the virial line in Fig. \ref{Fig:DFTeecorr1} determines the second virial coefficient $\rho_2(T)$. 
The $e-e$ corrected value is closer to the high-temperature benchmark 0.4917, but remains different.

\begin{table}[htp]
\caption{Virial representation of the dc conductivity $\sigma$ and of 
 $\rho_1^{\rm eff}(T,x)$ with $x=1/\ln(\Theta/\Gamma)$: the values for $\sigma^{\rm DFT-MD}$ 
 and $\rho_1^{\rm eff, DFT-MD}$ result from  DFT-MD simulations \cite{RSRB21}. The $e-e$ corrected values for $\sigma^{\rm ee-corr}$, 
 Eqs.~(\ref{reduction}), (\ref{eq:ReeFit}), and the corresponding $\rho_1^{\rm eff, ee-corr}$, Eq.~(\ref{rho1eff}), are also shown.}
\begin{center}
\begin{tabular}{|c|c|c|c|c|c|c|c|c|}
\hline
 $n$ & $k_BT$ & $\Gamma$ & $\Theta$& $1/\ln(\Theta/\Gamma)$ & $\sigma^{\rm DFT-MD}$ &$ \rho_1^{\rm eff, DFT-MD}$
 & $\sigma^{\rm ee-corr}$ &$\rho_1^{\rm eff, ee-corr}$  \\
{[}g/cm$^3$] &[eV] &  & &  & [$10^6/\Omega$ m]  & & [$10^6/\Omega$ m]  &\\
\hline
 2  & 75 	& 0.3285 		& 1.8257 	&0.58302		&  11.44  & 1.073   	&10.034	&1.2234\\
 2  & 100  	& 0.24637 	& 2.4343 	&0.43657 		&  15.26  & 0.9269 	&12.434	&1.1375\\ 
 3  & 100 	& 0.28203 	& 1.8577 	&0.53047 		&  16.85  & 1.020  	&14.412	&1.1925\\ 
 3  & 150  	&0.18802 		&2.7866 	&0.37092 		&  25.67  & 0.8603 	&20.157	&1.0956\\ 
 4  & 150 	&0.20694 		&2.3003 	&0.41522 		&  27.39  & 0.9026 	&22.058	&1.1208\\ 
\hline
\end{tabular}
\end{center}
\label{tab:Max1}
\end{table}

The $e-e$ correction factor (\ref{eq:ReeFit}) was obtained from a two-moment approximation. The inclusion of higher moments of the distribution function would change the $e-e$ correction factor and the corresponding values of $\rho_1,\rho_2(T)$. These corrections are expected to be small, see the good convergence with increasing number of moments reported in Ref. \cite{Redmer97}. 

The virial coefficient $\rho_2(T)$ is determined by the screening of the Coulomb potential. The Born approximation can be applied in the high temperature range $T_{\rm Ha} \gg 1$ so that strong collisions leading to a T-matrix approach are not relevant. 
A reason for the $\rho_2$ discrepancy can also be found in the treatment of screening, see the analysis of the slope parameter in \cite{SCCS22}.  
The electrons should be treated as dynamic screening, in lowest approximation by the RPA expression for the polarization function, to obtain the QLB benchmark (\ref{virLB}).
It is not clear to what extent the dynamic screening by electrons is already implemented in the DFT-MD simulations by the choice of the exchange-correlation part $E_{xc}$ of the energy-density functional. 

If $\sigma^{\rm Lorentz}(T,n)$ which is required for the calculation of $R_{ee}$, Eq. (\ref{eq:ReeFit}), treats the dynamical screening of the electron-ion interaction differently than the DFT-MD approach $\sigma^{\rm DFT-MD}(T,n)$, the use of the correction factor is not consistent since the slope parameter $\rho_2(T)$ is not corrected.
With respect to the contribution of the ions to the screening, the ion structure factor describes the  static screening which is well approximated in the DFT-MD simulations.

In conclusion, the correction factor (\ref{eq:ReeFit}) given in \cite{Heidi} reproduces the general behavior  for $\Theta \le 1$ on $R_{ee}\approx 1$  and is a reasonable approximation to the correct Spitzer result in the low-density limit. Only a two-moment  approximation of the electron distribution function is considered to calculate $R_{ee}$  (\ref{eq:ReeFit}). It can be improved by considering higher moments of the electron distribution function so that the exact value of the Spitzer benchmark is reproduced. 
With respect to the second virial coefficient, the specific choice of the $e-i$ interaction, which defines the Lorentz model plasma as a reference system, is discussed in the following Sec. \ref{Sec:syst}.

\subsection{Systematic approaches}
\label{Sec:syst}

While  electron-ion collisions are well understood to calculate the conductivity of plasmas, the inclusion of electron-electron collisions is notoriously difficult. One reason for this is that a relaxation time approach is only valid for elastic collisions where the single-particle energy is conserved, what holds to a good approximation for $e - i$ collisions, but not for $e-e$ collisions.
Several approaches have been worked out to implement $e-e$ collisions. 
The kinetic theory can solve this problem starting from the linearized Boltzmann equation. The perturbation of the single electron distribution function in a weak electrical field  results from the Kohler variational principle.
In dense plasmas, gLRT is used to express the conductivity in terms of equilibrium correlation functions \cite{Spitzer53,Roep88}.

To calculate the conductivity of degenerate electrons in condensed matter, the Ziman formula is known, which considers the force-force autocorrelation function. 
Calculations of the collision frequency  based on the solution of the electron-ion scattering problem are improved by a renormalization factor $r(\omega)$ 
to obtain the correct dynamic collision frequency  of the plasma \cite{Reinholz00,Reinholz05,Heidi}.
While the Ziman formula is frequently used to describe degenerate Coulomb plasmas such as electrons in liquid metals or solids, the renormalization factor allows to  treat also non-degenerate plasmas, especially in the low-density range. The method of including $e-e$ correlations by means of a renormalization factor $r(\omega)$ is equivalent to using the correction factor $R_{ee}$ and requires analogous approximations.

An improved approach to treat dense plasmas is to start from the Lorentz model plasma, in which the $e-i$ interaction is solved, and the electron distribution function is rigorously obtained from the relaxation time ansatz for any degree of degeneracy.
Starting from the linearized Boltzmann equation, thermoelectric transport coefficients for Lorentz model plasmas are calculated, see the Lee-More model \cite{LeeMore}.
For a dense system, 
the averaged atom model was worked out,  which describes the average influence of the surrounding plasma on the atom.
The interaction of the electrons with the ion system is better described by the DFT approach, and the transition to condensed matter is well established. The electron-electron interaction is only considered in mean-field approximation to introduce an effective potential for the single-electron quasiparticle states. 

The success of relaxation-time models for transport coefficients to treat the electron-ion interaction  has led to develop a generalization of these models which includes $e-e$ collisions. 
For example, as in Refs. \cite{Starrett20,Starrett20a}, the relaxation-time model by Starrett \cite{Starrett17} includes
Pauli blocking and accounts for correlations using a mean-
force scattering cross-section for electron-ion collisions, but
electron-electron collisions are accounted for only through the
correction formula for $R_{ee}$ proposed in Ref. \cite{Heidi}. 

A possible solution for this particular problem  of including $e-e$ collisions  in the determination of plasma conductivity can be the use of path-integral Monte Carlo (PIMC) simulations to calculate $\sigma(T,n)$. 
In PIMC simulations,  $e-e$ correlations are treated in an exact way.
This enables  the immediate calculation of $R_{ee}(T,\Theta)$ according to (\ref{reduction}) and the construction of consistent interpolation formulas. 
In particular, the determination of virial coefficients would be of interest.
Such PIMC simulations are not yet possible. 
Highly accurate results are only available for the uniform electron gas \cite{RDVBB24}.
In addition, there is the unsolved sign problem, and PIMC simulations are currently performed for too small numbers of particles (a few tens). 
It is expected that significant results for the conductivity of hydrogen plasmas from PIMC simulations will be available in the near future.
This will overcome the drawback of DFT-MD simulations where $e-e$ correlations are not treated precisely.

We propose to use the virial expansion (\ref{eq:virial}) for the determination of the conductivity of the hydrogen plasma  in order to 
analyze the consistency of different theoretical approaches using the virial coefficients as a benchmark.
Having $\rho_1, \rho_2(T),$ etc. available, we can find improved versions of the correction factor $R_{ee}(T,n)$.
Since these correction factors relate the plasma conductivity $\sigma(T,n)$ to the conductivity of a Lorentz model plasma $\sigma^{\rm Lorentz}(T,n)$, see Eq. (\ref{reduction}),
these correction factors depend on the definition of the Lorentz model Hamiltonian. 
They depend in particular on the definition of the effective electron-ion potential, which includes dynamic or static screening by electrons, or on the particular choice of the exchange-correlation energy $E_{xc}$ in DFT-MD calculations.
If the determination of the conductivity and virial expansion for the Lorentz model plasma is known, a corresponding correction factor can be defined.

In particular, we consider the correction factor 
\begin{equation}
\label{reductionDFT}
R_{ee}^{\rm DFT-MD}(T,n)=\frac{\sigma(T,n)}{\sigma^{\rm ^{\rm DFT-MD}}(T,n)}
\end{equation}
and use the virial expansions (\ref{VirH}), (\ref{VirDFT}).
The virial expansion for the high-temperature limit results in
\begin{equation}
\lim_{T \to \infty}R_{ee}^{\rm DFT-MD}(T,x) = 0.5814+0.8304 \,x
\label{reductionDFTv}
\end{equation}
in contrast to (\ref{Reevir}) which describes the relationship between the conductivity of a model plasma and a Lorentz model plasma with the same form of screened interaction.
The expression (\ref{reductionDFTv}) is the correction factor required to obtain the conductivity of hydrogen plasmas in the low-density, high-temperature  limit after performing DFT-MD simulations. 
Compared to (\ref{Reevir}), which is determined in the framework of a two-moment approximation, the first number 0.5814 results from the consideration of higher moments of the distribution function. The second number 0.8304 results from the treatment of the screening in the special version of the DFT approach and cannot be analyzed in this work.
The advantage of the correction factor $R_{ee}(T,\Theta)$ is that its behavior is known in both limits: for $\Theta \le 1$ it is close to 1, and for $\Theta \gg 1$ it is determined by the virial expansion.
Note that the correction factor $R_{ee}(T,n,Z)$ in \cite{Heidi} is given for any ionic charge $Z$, so that it can also be applied to other substances. 

\section{Conclusions}
\label{sec:Concl}

We propose an exact virial expansion (\ref{eq:virial}) for the plasma conductivity to analyze 
the consistency of theoretical approaches. 
The virial coefficients serve as a benchmark for different approaches and can be better visualized using virial plots. 
For example, various analytical calculations of 
the dc conductivity $\sigma(T,n)$, which were presented in Ref.~\cite{Grabowski} do not fulfill this exact 
requirement and do not provide accurate results in the range of small densities. 
The results of DFT-MD simulations, which are currently considered to be the most reliable, are checked by virial plots, and shortcomings are found in the region of small densities.
By benchmarking with the virial expansion (\ref{eq:virial}) for $x \to 0$, future PIMC simulations 
 can also be tested.
 These \textit{ab initio} 
simulations become a computational challenge in the low-density region, but the virial 
expansion enables extrapolation into this range. For the construction of interpolation 
formulas, see~\cite{Esser03}, knowledge of the virial coefficients is an important ingredient.

A particular problem concerns the treatment of the $e-e$ interaction in DFT-MD simulations. 
To a certain extent, the $e-e$ interaction is included  in the exchange-correlation energy $E_{xc}$ in order to obtain an effective mean field for optimal single-particle states. 
However, with the help of virial expansion it was shown that $e-e$ collisions are not included in the DFT-MD calculations. 
By introducing a correction factor $R_{ee}$, which is obtained from a model plasma with statically screened Coulomb interaction, the first virial coefficient  $\rho_1$ can be reproduced. 
In this work, the problems with the second virial coefficient $\rho_2(T)$, which is determined by the treatment of dynamic screening, are emphasized. 
Since the dynamic screening is not considered in the DFT-MD simulations, the second virial coefficient  $\rho_2^{\rm QLB}$ (\ref{virLB}) is also not obtained by the correction factor $R_{ee}$. 
A numerical calculation can be considered in the context of PIMC simulations to determine the second virial coefficient $\rho_2(T)$. 

A correction factor $R_{ee}^{\rm DFT-MD}(T,n)$ was introduced which takes into account the specific choice of exchange-correlation energy in the DFT approach. 
The high-temperature limit of the virial expansion was given in Eq. (\ref{reductionDFTv}), and the full $T$ dependence would be of interest. 
 The DFT-MD simulations are very accurate at higher densities since $e-e$ correlations are strongly suppressed due to Pauli blocking in the degenerate range ($\Theta \le 1$). 
 With the knowledge of $R_{ee}^{\rm DFT-MD}(T,n)$, results for the conductivity of hydrogen plasmas can also  be obtained from DFT-MD calculations for the non-degenerate range.
In addition to the analytical evaluation of correlation functions for higher moments of the distribution function in the framework of gLRT, 
the numerical calculation in the context of PIMC simulations can be considered to determine the second virial coefficient $\rho_2(T)$. 

The approach described here can also be applied to other transport properties such as thermal conductivity, thermopower, viscosity and diffusion coefficients \cite{French22}. 
It is also interesting to extend the virial expansion to substances other than hydrogen, where different ions can be formed and the electron-ion interaction is no longer purely Coulombic.\\

{\bf Acknowledgments}\\

The author would like to thank M. Bethkenhagen, M.~French, R. Redmer, H. Reinholz and M. Sch\"orner for valuable and fruitful discussions.


\end{document}